# High Temperature Enthalpy Increment and Thermodynamic Functions of ZrCo: An Experimental and Theoretical Study

D. Chattaraj[*], Ram Avtar Jat, S.C. Parida, Renu Agarwal, Smruti Dash

Product Development Division, Bhabha Atomic Research Centre,

Trombay, Mumbai – 400 085, India

[*]*Corresponding Author*: Debabrata Chattaraj

*Postal Address*:

Product Development Division,

Bhabha Atomic Research Centre,

Trombay, Mumbai – 400 085,

India

Tel. No. - +91 22 2559 4805

Fax: +91 22 2550 5151

*E-mail*: debchem@barc.gov.in

**Abstract**

The ZrCo intermetallic was proposed as tritium storage material in the International Thermonuclear Experimental Reactors (ITER) project. The thermodynamic properties of ZrCo intermetallic were investigated both experimentally and theoretically to determine its applicability for the storage of hydrogen isotopes. The enthalpy increments of ZrCo were measured using a high temperature inverse drop calorimeter in the temperature range 645-1500 K. A set of thermodynamic functions such as entropy, Gibbs energy function, heat capacity, Gibbs energy and enthalpy values for ZrCo were calculated using the data obtained in this study. The polynomial expression of enthalpy increments and heat capacity obtained for ZrCo(s) in the temperature range 642-1497 K are given as:

$$H_m^0(T) - H_m^0(298.15\,K)\,(J \cdot mol^{-1}) = 25.682 \cdot (T/K) + 29.807 \cdot 10^{-4} \cdot (T/K)^2 + 2.1864 \cdot 10^5 \cdot (K/T) - 8655.5$$

$$C_{p.m}^0\,(J \cdot K^{-1} \cdot mol^{-1}) = 25.682 + 5.961 \cdot 10^{-3} \cdot (T/K) - 2.1864 \cdot 10^5 \cdot (K/T)^2$$

The enthalpy of formation of ZrCo at 0 K was calculated as -55 kJ mol$^{-1}$ using the *ab-initio* method. The entropy ($S$) and specific heat capacities ($C_v$ and $C_p$) of ZrCo were also computed using Debye-Grüneisen quasi-harmonic approximation. A good agreement between the experimental and theoretically calculated values of specific heat ($C_P$) and entropy was obtained.

# 1. Introduction

Metal hydrides have drawn considerable attention as solid state materials for the storage of hydrogen isotopes [1]. Though uranium is conventionally used as getter bed for hydrogen isotopes, because of its pyrophoricity the search for an alternate material is important. As an alternate, ZrCo intermetallic has been proposed as storage material for hydrogen isotopes in the fusion reactor project because of its favorable properties [2]. ZrCo absorbs hydrogen at room temperature to form the hydride with a maximum stoichiometry of $ZrCoH_3$. The hydrogen equilibrium pressure of ZrCo hydride is around $10^{-3}$ Pa at room temperature and ~$10^5$ Pa at 673 K [3]. However, the major drawback of ZrCo is due to its poor absorption-desorption cycling property during prolonged thermal cycling [4] due to the disproportionation of its hydride ($ZrCoH_x$ (x ≤ 3)) into the stable hydride phase $ZrH_2$ and the hydrogen non-absorbing phase $ZrCo_2$.

There is plenty of experimental work on the thermal properties of ZrCo and $ZrCoH_3$ system whereas few theoretical reports are available. Gachon etal. measured the enthalpy of formation of different ZrCo phases like $Co_{0.33}Zr_{0.67}$, $Co_{0.5}Zr_{0.5}$, $Co_{0.67}Zr_{0.33}$, $Co_{0.80}Zr_{0.20}$ by direct reaction calorimetry at high temperatures in the neighbourhood of the phase melting point [5]. They have reported the enthalpy of formation of ZrCo intermetallic as -42.2 ± 1 kJ/mol at 1512 K. The stability of ternary hydrides with respect to their decomposed state (i.e. to a transition metal intermetallic compound and hydrogen gas) was investigated in terms enthalpy of formation by Miedema etal [6]. Kuniaki etal.[7] experimentally investigated the stability of ZrCo and ZrNi to heat cycles in hydrogen atmosphere. The reversible disproportionation of ZrCo under high temperature and hydrogen pressure was studied by Konishi etal. [8]. Bekris etal. investigated the rate of hydrogen induced disproportionation of ZrCo to a broder temperature range [9]. The thermodynamic properties of ZrCo like Helmholtz free energy $F$, the internal energy $E$, entropy $S$ and constant-volume specific heat $C_V$ are determined with the harmonic approximation by Gan Li eta [10]l. The calculated thermodynamic data of Zr was compared with the experimental results. The thermodynamic functions of ZrCo were calculated within the temperature 300 – 1000 K using density-functional perturbation theory. In our previous computational study, we reported the enthalpy of formation of ZrCo and $ZrCoH_3$ at 0 K [11]. The enthalpy of formation of ZrCo and $ZrCoH_3$ at 0 K are found to be -56 kJ/mol and -91 kJ/mol of $H_2$, respectively without the zero point energy contribution. After including the zero point energy, the enthalpy of formation of ZrCo intermetallic at 0 K is found to be -51 kJ/mol. The coefficient of electronic specific heat (γ) calculated in our study were found to be -6.505 mJ/mol-$K^2$ and 7.021mJ/mol-$K^2$ for ZrCo and $ZrCoH_3$ respectively.

To know the behavior of ZrCo under different thermal conditions, it is necessary to study the thermodynamic properties of this intermetallic. For this purpose, here, the enthalpy increment and other thermodynamic functions for ZrCo have been studied experimentally. The *ab-initio* calculations of ZrCo have also been performed to support the experimental findings. This study will help to explore the basic thermal properties of ZrCo, in future, which will hopefully help to overcome its drawbacks related to tritium storage.

## 2. Experimental Details

### 2.1 Synthesis and characterization

ZrCo intermetallic was prepared by arc-melting stoichiometric mixtures of high purity zirconium (purity 99.98%) and cobalt (purity 99.95%) metals. The cast ingot was wrapped in a titanium foil and annealed in an evacuated sealed quartz tube at 973 K for 72 hrs and water quenched to room temperature. The annealed sample was cut into pieces of 4 mm x 2 mm x 1 mm using a diamond saw cutter for carrying out the calorimetric studies. The polished specimen of synthesized intermetallic was characterized by X-ray powder diffraction method on a STOE X-ray diffractometer. The X-ray diffraction pattern of ZrCo was shown in Fig. 1. The formation of ZrCo phase (cubic structure) was confirmed by comparing XRD pattern obtained in this study with the reported in the JCPDS file no: 18-0436.

### 2.2 Inverse drop calorimetric measurement

The heat contents corresponding to the enthalpy increments from ambient temperature to the respective temperature of a given run were determined using SETARAM multi-detector high temperature calorimeter (MHTC-96) operating in drop mode. All measurements were performed in flowing high purity argon.

From the reference material drops, the actual instrument sensitivity can be determined as:

$$S = \frac{\int(\varphi_R - \varphi_B)d\tau}{\int_{T_a}^{T_m} C_{P,R}^0 dT} \cdot \frac{M_R}{m_R} \quad (1)$$

where $\phi_R$, $\phi_B$, $T_a$ and $T_m$ are measured heat of reference and blank, and the ambient and measurement temperature, respectively. The $m_R$ and $M_R$ are the reference material mass and molar mass, respectively, $C_{P,R}^0$ stands for molar heat capacity of the reference material. The enthalpy increment corresponding to heating the sample material from $T_a$ to $T_m$ is given by:

$$H_m^0(T) - H_m^0(298.15\ K) = \int \frac{(\varphi_S - \varphi_B)d\tau}{S} \cdot \frac{M_S}{m_S} \tag{2}$$

where $\phi_s$ is the measured heat of the sample.

The calorimeter was calibrated by dropping samples of synthetic sapphire (SRM 720) supplied by NIST, USA. In a typical experiment at a given temperature, the ZrCo and $Al_2O_3$(s), maintained at the ambient temperature were dropped, alternatively, into the calorimeter maintained at required temperature.

### 2.3 Computational Details

All calculations were performed using Vienna *ab initio* simulation package (VASP) [12,13], which implements Density Functional Theory (DFT) [14]. The plane wave based pseudo-potential method has been used for the total energy calculations. The electron-ion interaction and the exchange correlation energy were described under the Projector-augmented wave (PAW) [15, 16] method and the generalized gradient approximation (GGA) of Perdew–Burke–Ernzerhof (PBE) [17] scheme, respectively. All-electronic projector-augmented wave potentials were employed for the elements Zr and Co. The valence electron configurations of Zr and Co were set to $5s^1 4d^3$, $4s^1 3d^8$, respectively. The energy cut off for the plane wave basis set was fixed at 500 eV. Total energies for each relaxed structure using the linear tetrahedron method with Blöchl corrections were subsequently calculated in order to eliminate any broadening-related uncertainty in the energies [18]. Ground state atomic geometries were obtained by minimizing the Hellman-Feynman forces [19,20] using the conjugate gradient method. The force on each ion was minimized upto 5 meV/Å. In order to verify the magnetic nature of the systems studied in this work, we have performed the total energy calculations using the spin polarized version of the DFT. The k-point meshes were constructed using the Monkhorst-Pack scheme [21] and the 9x9x9 k-point mesh was used for the primitive cell for Brillouin zone sampling.

The quasi-harmonic Debye model was used to calculate the thermodynamic functions of the compounds [22-24]. The non-equilibrium Gibbs function $G^*(V,P,T)$ can be estimated as:

$$G^*(V,P,T) = E(V) + PV + A_{vib}[\Theta(V);T] \tag{3}$$

Where $E(V)$ is the total energy per unit cell of the compounds, $PV$ is the constant hydrostatic pressure condition, $\Theta(V)$ is the Debye temperature and $A_{vib}$ is the Helmholtz free energy which can be expressed as

$$A_{vib}(\Theta,T) = nkT\left[\frac{9\Theta}{8T} + 3\ln\left(1 - e^{-\Theta/T}\right) - D\left(\Theta/T\right)\right] \qquad (4)$$

where $n$ is the number of atoms per formula unit, $D(\Theta/T)$ describe the Debye integral. The Debye temperature $\Theta$ is

$$\Theta = \frac{\hbar}{k}\left[6\pi^2 V^{1/2} n\right]^{1/3} f(\sigma)\left(\frac{B_S}{M}\right)^{1/2} \qquad (5)$$

where $M$ is the molar mass per unit cell and $B_S$ is the adiabatic bulk modulus, which is approximated by the static compressibility

$$B_S \approx B(V) = V\left(\frac{d^2 E(V)}{dV^2}\right), \qquad (6)$$

$f(\sigma)$ is given by Eq. (5) [36,37]

$$f(\sigma) = \left\{3\left[2\left(\frac{21+\sigma}{31-2\sigma}\right)^{3/2} + \left(\frac{11+\sigma}{31-\sigma}\right)^{3/2}\right]^{-1}\right\}^{1/3} \qquad (7)$$

where $\sigma$ is poisson ratio. Therefore, the non-equilibrium Gibbs function $G^*(V,P,T)$ can be minimized as a function of volume $V$ as

$$\left[\frac{\partial G^*(V,P,T)}{\partial V}\right]_{P,T} = 0 \qquad (8)$$

The thermal equation-of-state (EOS) $V(P,T)$ can be obtained by solving the eq. (6). The isothermal bulk modulus $B_T$ is given by

$$B_T(P,T) = V\left(\frac{\partial^2 G^*(V,P,T)}{\partial V^2}\right)_{P,T} \qquad (9)$$

The thermodynamic quantities, e.g., heat capacities $C_V$ at constant volume and $C_P$ at constant pressure, and entropy $S$ have been calculated by applying the following relations:

$$C_V = 3nk\left[4D(\Theta/T) - \frac{3\Theta/T}{e^{\Theta/T}-1}\right], \tag{10}$$

$$C_P = C_V(1 + \alpha\gamma T), \tag{11}$$

$$S = nk\left[4D(\Theta/T) - 3\ln\left(1 - e^{-\Theta/T}\right)\right] \tag{12}$$

where α is the thermal expansion coefficient and γ is the Grüneissen parameter which are given by following equations:

$$\alpha = \frac{\gamma C_V}{B_T V}, \tag{13}$$

$$\gamma = -\frac{d\ln\Theta(V)}{d\ln V}\pi r^2 \tag{14}$$

## 3. Results and discussion

The experimentally determined enthalpy increment data acquired by drop mode of multi-HTC are given in table 1 and shown in Fig. 2. Enthalpy increment values were least squared fitted in the temperature rang 642-1,497 K (before melting) into polynomial equation using Shomate method [7] with constraints, i) $H_m^0(T) - H_m^0(298.15\,K) = 0$ at 298.15 K and ii) $C_{p.m}^0(298.15\,K) = 25.0$ J K$^{-1}$mol$^{-1}$. $C_{p.m}^0(298.15\,K)$ value was estimated from the heat capacity of component elements. The polynomial expression obtained for ZrCo(s) in the temperature range 642-1497 K is given as:

$$H_m^0(T) - H_m^0(298.15\,K)(J \cdot mol^{-1}) = 25.682 \cdot (T/K) + 29.807 \cdot 10^{-4} \cdot (T/K)^2 +$$
$$2.1864 \cdot 10^5 \cdot (K/T) - 8655.5 \tag{15}$$

The heat capacity expression of ZrCo(s) was obtained by differentiating above enthalpy increment expressions with respect to temperature. The heat capacity expression obtained from Multi HTC data can be given as:

$$C_{p.m}^0(J \cdot K^{-1} \cdot mol^{-1}) = 25.682 + 5.961 \cdot 10^{-3} \cdot (T/K) - 2.1864 \cdot 10^5 \cdot (K/T)^2 \quad (642 \leq T/K \leq 1497) \tag{16}$$

Enthalpy increment values for ZrCo (s) have not been previously reported in the literature to compare. In order to generate thermodynamic functions for ZrCo (s) from enthalpy increment data; $\Delta_f H_m^0(298.15\ K)$ and $S_m^0(298.15\ K)$ values are required. Here, using the first principle method, we have computed $\Delta_f H_m^0(ZrCo, s, 0\ K) = -55\ kJmol^{-1}$ and $S_m^0(ZrCo, s, 298.15\ K) = 37.2\ J \cdot K^{-1} \cdot mol^{-1}$. This value is selected in this study for the calculations of thermodynamic functions of ZrCo.

The thermodynamic functions like $S_m^0(T)$, $C_{P,m}^0(T)$, $\{H_m^0(T) - H_m^0(298.15\ K)\}$, $\varphi_m^0(T)\{-(G_m^0(T) - H_m^0(298.15))/K\}$ have been calculated in the temperature range 298.15-1500 K using $\Delta_f H_m^0(298.15\ K) \approx$ -55 kJ/mol , $S_m^0(ZrCo, s, 298.15\ K) = 37.2\ J \cdot K^{-1} \cdot mol^{-1}$ and heat capacity data of ZrCo (s). The calculated thermodynamic functions are given in table 2.

The heat capacities ($C_v$ and $C_p$) and entropy (S) of ZrCo have been obtained from the *ab-initio* calculations are shown in Fig. 3 and Fig. 4, respectively along with the experimental data. It is evident from Fig. 3 that the theoretically calculated $C_p$ and $C_v$ of ZrCo are very close to each other upto 450 K, beyond which $C_p$ is higher than $C_v$ as is expected due to lattice thermal expansion. It is also observed that the experimentally determined $C_p$ is higher than that of theoretically calculated value. The reason for this discrepancy can be attributed to the additional contributions arising from the anharmonicity effect and electron phonon coupling which are not taken into account in the theoretical calculation.

## 4. Conclusions

The enthalpy increments of ZrCo (s) intermetallic were measured using drop calorimeter. The thermodynamic functions of ZrCo (s) were estimated using enthalpy increment data obtained in this study. The Gibbs energy values calculated for ZrCo (s) could be used for calculation of equilibrium pressure of hydrogen isotopes in hydriding and dehydriding cycles. The thermodynamic functions obtained experimentally and theoretically would be very much useful for the understanding of various thermal reactions of ZrCo intermetallic.

**Acknowledgements**

The authors are thankful to Dr. S.K. Mukerjee, Head, PDD and Dr. K.L. Ramakumar, Director, Radiochemistry and Isotope Group, for their support and encouragement in this study.

**Table captions**

Table 1 Enthalpy increment data along with fit values of ZrCo(s).

Table 2 Thermodynamic functions of ZrCo (s).

Table 1 Enthalpy increment data along with fit values of ZrCo(s).

| T/K | $H^o_m(T)-H^o_m(298.15\ K)$ $J\ mol^{-1}$ Experimental | $H^o_m(T)-H^o_m(298.15\ K)$ $J\ mol^{-1}$ Calculated | % difference[*] |
|---|---|---|---|
| 642.0 | 9369 | 9401 | -0.33 |
| 692.9 | 10809 | 10886 | -0.69 |
| 743.6 | 12304 | 12384 | -0.63 |
| 794.3 | 13858 | 13900 | -0.27 |
| 844.0 | 15409 | 15402 | 0.07 |
| 893.9 | 16990 | 16928 | 0.39 |
| 943.5 | 18570 | 18461 | 0.62 |
| 994.0 | 20196 | 20037 | 0.82 |
| 1043.8 | 21764 | 21608 | 0.75 |
| 1095.9 | 23399 | 23269 | 0.72 |
| 1144.3 | 24991 | 24826 | 0.66 |
| 1194.8 | 26555 | 26467 | 0.33 |
| 1245.1 | 28227 | 28118 | 0.39 |
| 1295.3 | 29772 | 29780 | -0.03 |
| 1345.3 | 31408 | 31452 | -0.14 |
| 1397.0 | 32980 | 33196 | -0.65 |
| 1446.8 | 34645 | 34892 | -0.71 |
| 1497.2 | 36371 | 36623 | -0.69 |

[*]% difference = {(Exp.-Cal.)./Exp}·100

Table 2 Thermodynamic functions of ZrCo (s).

| $T$/K | $C_{p,m}^0(T)$/ J·K$^{-1}$mol$^{-1}$ | $H_m^0(T) - H_m^0(298.15\ K)$/ kJ·mol$^{-1}$ | $S_m^0(T)$/ J·K$^{-1}$mol$^{-1}$ | $\varphi_m^0(T)$/ J·K$^{-1}$mol$^{-1}$ |
|---|---|---|---|---|
| 298.15 | 24.11 | 0.0 | 37.2 | 37.2 |
| 300 | 24.15 | 0.04 | 37.3 | 37.2 |
| 400 | 25.51 | 2.53 | 44.5 | 38.2 |
| 500 | 26.30 | 5.13 | 50.3 | 40.0 |
| 600 | 26.86 | 7.79 | 55.1 | 42.1 |
| 700 | 27.32 | 10.50 | 59.3 | 44.3 |
| 800 | 27.72 | 13.25 | 63.0 | 46.4 |
| 900 | 28.09 | 16.04 | 66.3 | 48.5 |
| 1000 | 28.44 | 18.87 | 69.3 | 50.4 |
| 1100 | 28.78 | 21.73 | 72.0 | 52.2 |
| 1200 | 29.11 | 24.62 | 74.5 | 54.0 |
| 1300 | 29.43 | 27.55 | 76.5 | 55.6 |
| 1400 | 29.74 | 30.51 | 79.0 | 57.2 |
| 1500 | 30.06 | 33.50 | 81.1 | 58.8 |

**Figure Captions**

Figure 1 XRD pattern of ZrCo (s)

Figure 2 Enthalpy increment of ZrCo (s)

Figure 3 Specific heat of ZrCo (s)

Figure 4 Entropy of ZrCo (s)

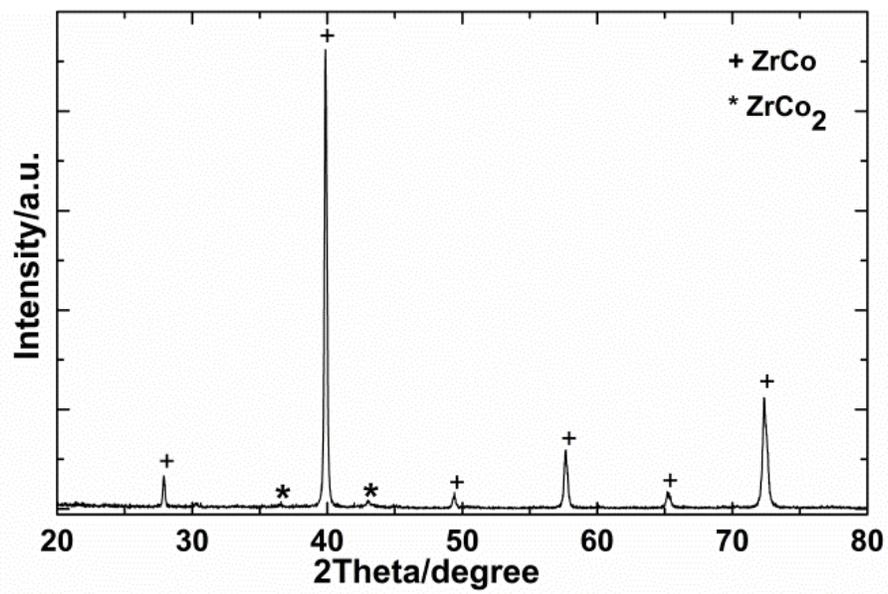

Fig. 1

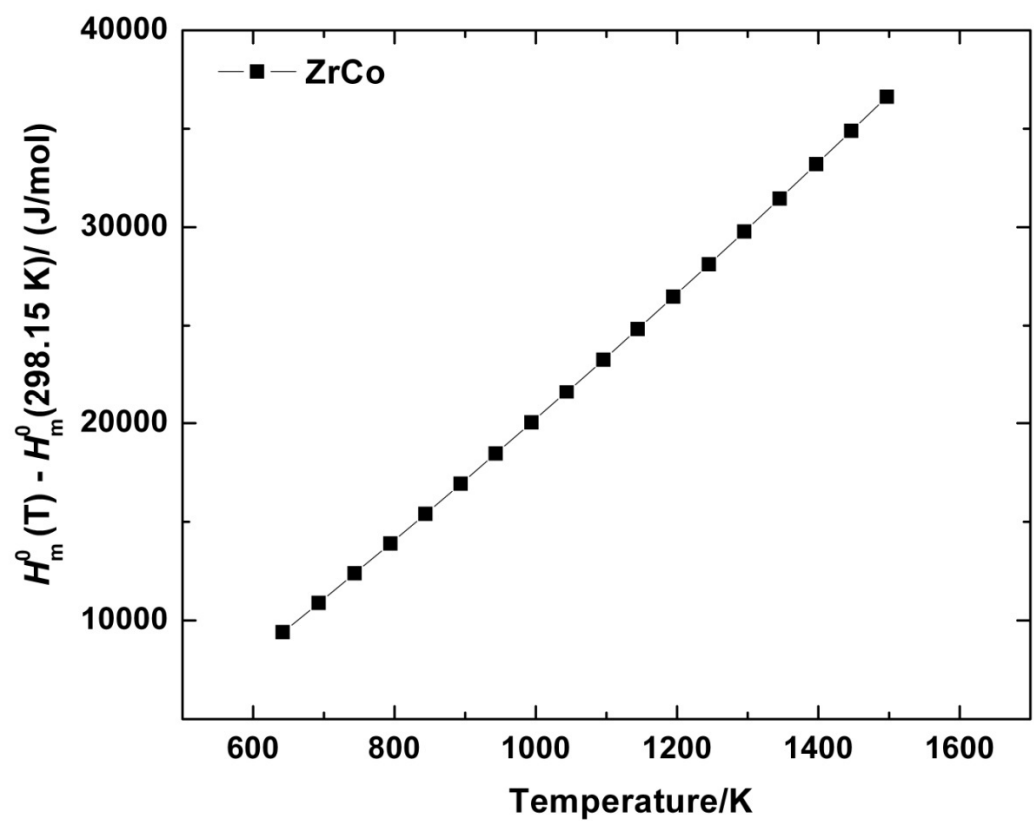

Fig. 2

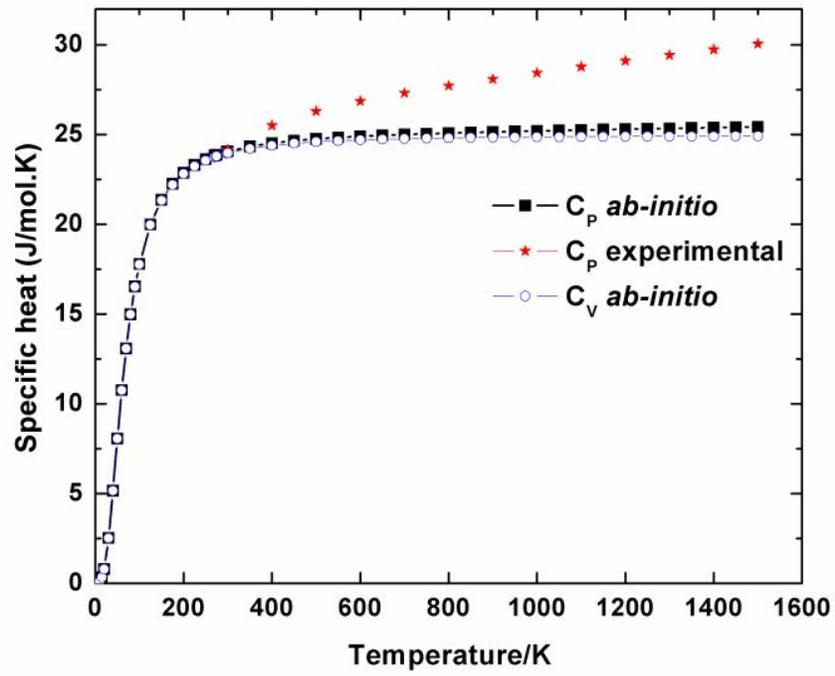

Fig. 3

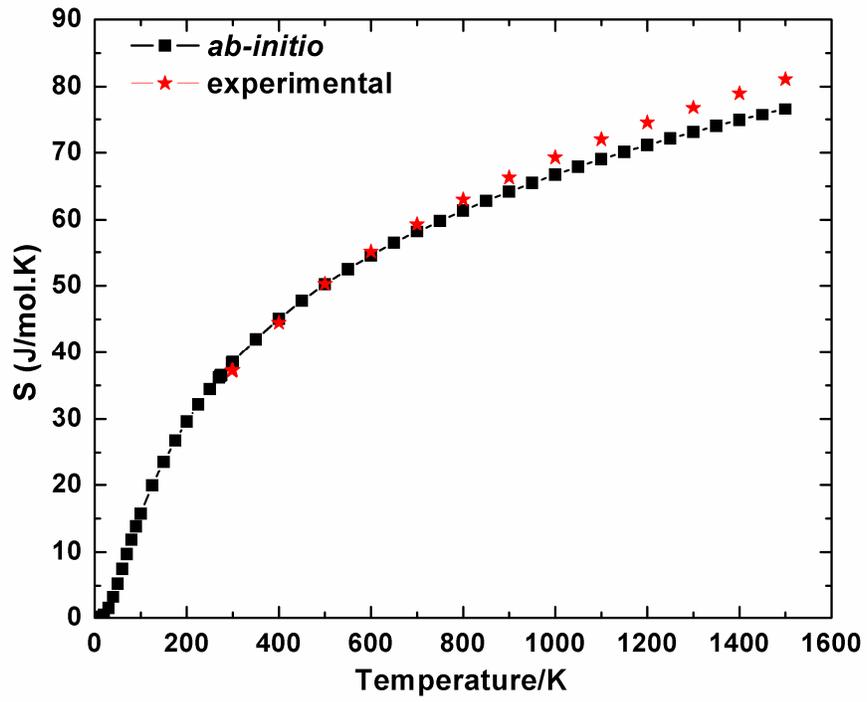

Fig. 4